\documentclass[aps,floats,superscriptaddress,showpacs]{revtex4}
\usepackage{amsmath}

\usepackage{epsfig}
\usepackage{graphicx}
\usepackage{dcolumn}
\usepackage{bm}

\def\gapp{\lower.35em\hbox{$\stackrel{\textstyle>}{\sim}$}}
\def\lapp{\lower.35em\hbox{$\stackrel{\textstyle<}{\sim}$}}

\begin{document}
\bibliographystyle{apsrev}

\title{Gauge fields and curvature in graphene}
\author{Mar\'{\i}a A. H. Vozmediano, Fernando de Juan and Alberto Cortijo}
\affiliation{Instituto de Ciencia de Materiales de Madrid,\\
CSIC, Cantoblanco, E-28049 Madrid, Spain.}

\date{\today}
\begin{abstract}
The low energy excitations of graphene can be described by a
massless Dirac equation in two spacial dimensions. Curved graphene
is proposed to be described by coupling the Dirac equation to the
corresponding curved space. This covariant formalism gives rise to
an effective hamiltonian with various extra terms. Some of them
can be put in direct correspondence with more standard tight
binding or elasticity models while others are more difficult to
grasp in standard condensed matter approaches. We discuss this
issue, propose models for singular and regular curvature and
describe the physical consequences of the various proposals.
\end{abstract}
%

\maketitle

\section{Introduction}

Since its experimental realization graphene has been a focus of
intense research activity both theoretical and experimentally as
can be seen in the recent reviews \cite{GN07,KN07,GSC07,RMP08}.
The origin of this interest lies partially on the experimental
capability to tailoring the samples into special geometries
leading to graphene based electronics but it is its unusual
electronic properties and the breakdown of the standard Fermi
liquid description  what has lead the main activity in the field.
Despite the intense research three aspects of the electronic
properties remain to be fully understood: that of the minimal
conductivity \cite{Netal05,ZTSK05,CJetal08}, the observed charge
inhomogeneities at low densities \cite{Y07}, and the exceptionally
high mobility at room temperature with an associated mean free
path of the order of the size of the samples
\cite{BSetal08,DSetal08}. A key point to understand all the three
problems is a better knowledge of the nature and effect of
disorder in graphene.

One of the most intriguing properties of the suspended graphene
samples is the observation of mesoscopic corrugations in both
suspended \cite{Metal07,BBetal08} and deposited on a substrate
\cite{Setal07,ICCFW07} samples whose possible influence on the
electronic properties only now starts to be realized
\cite{Metal07,Metal07b,CN07,JCV07,GKV07,BP08}. Although the
observed ripples were invoked from the very beginning to explain
the absence of weak localization in the samples
\cite{Metal06,MG06}, there have been so far few attempts to model
the corrugations. The two main approaches  are based either on the
presence of disclinations and other topological defects in the
graphene lattice \cite{CV07a,CV07b} or on the theory of elasticity
\cite{Metal07,Metal07b,CN07}. The concrete realization of both
models in the continuum gives rise to the appearance of gauge
fields coupled to the electronic degrees of freedom \cite{GBD08}.
In this work we will describe two different approaches to model
curvature in graphene and study its influence on the physical
properties of the material.

\section{A summary of graphene features}
Under a theoretical point of view the synthesis of graphene has
opened  a new world where ideas from different branches of physics
can be confronted and tested in the laboratory. On the electronic
point of view it can be shown that the low energy excitations of
the neutral system obey a massless Dirac equation in two
dimensions. This special behavior originates on the geometry and
topology of the honeycomb lattice and has profound implications to
the transport and optical properties. Although the Fermi velocity
is approximately a hundredth of the speed of light, the
masslessness of the quasiparticles brings the physics to the
domain of relativistic quantum mechanics where phenomena like the
Klein paradox or the Zitterbewegung \cite{KN07} can be explored.
None of these questions arise within the  quantum field theory
approach but  its full applicability to the condensed matter
system is questionable.
\begin{figure}
\begin{center}
\includegraphics[width=4.5cm]{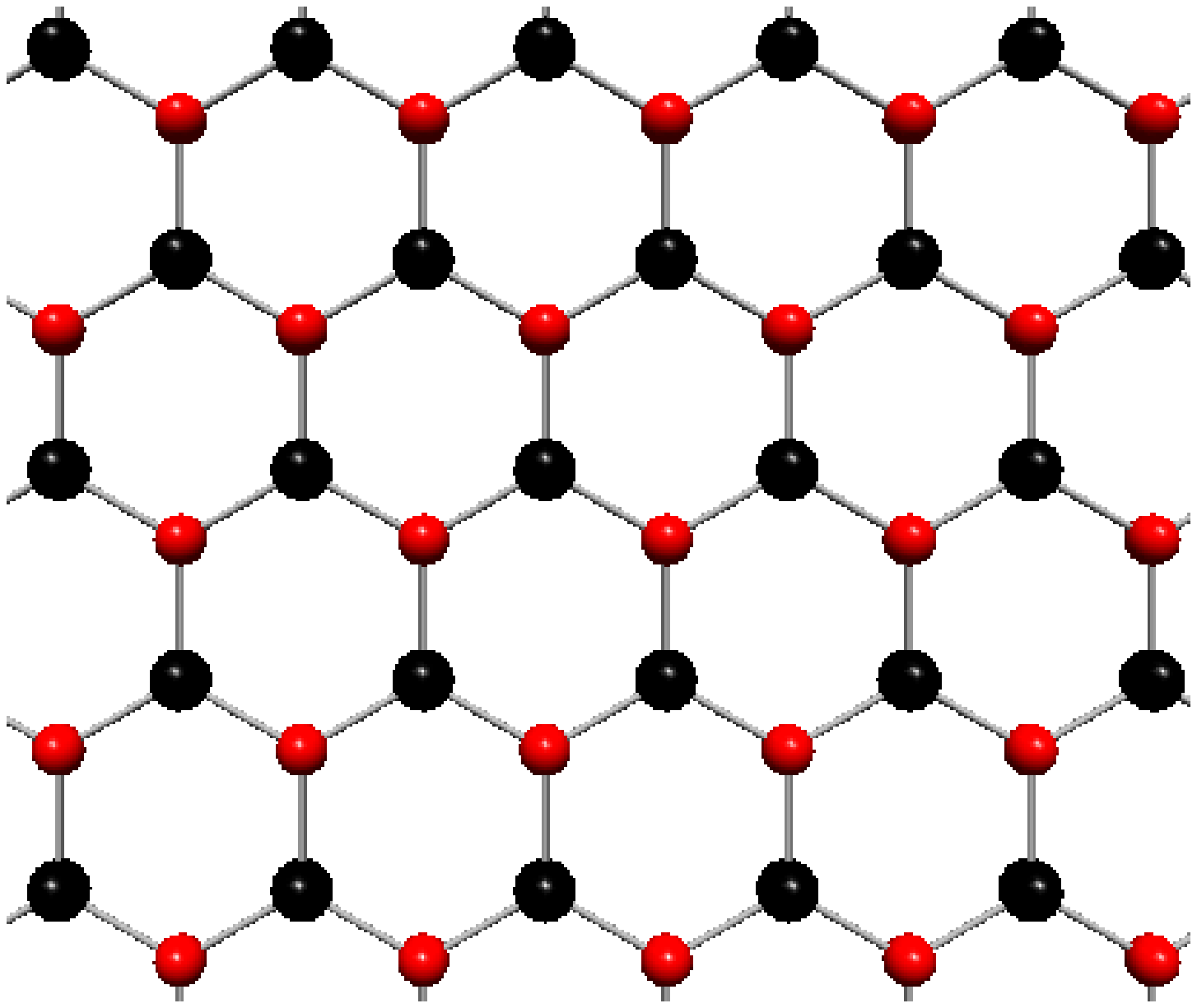}
\hspace{1cm}
\includegraphics[width=5.5cm]{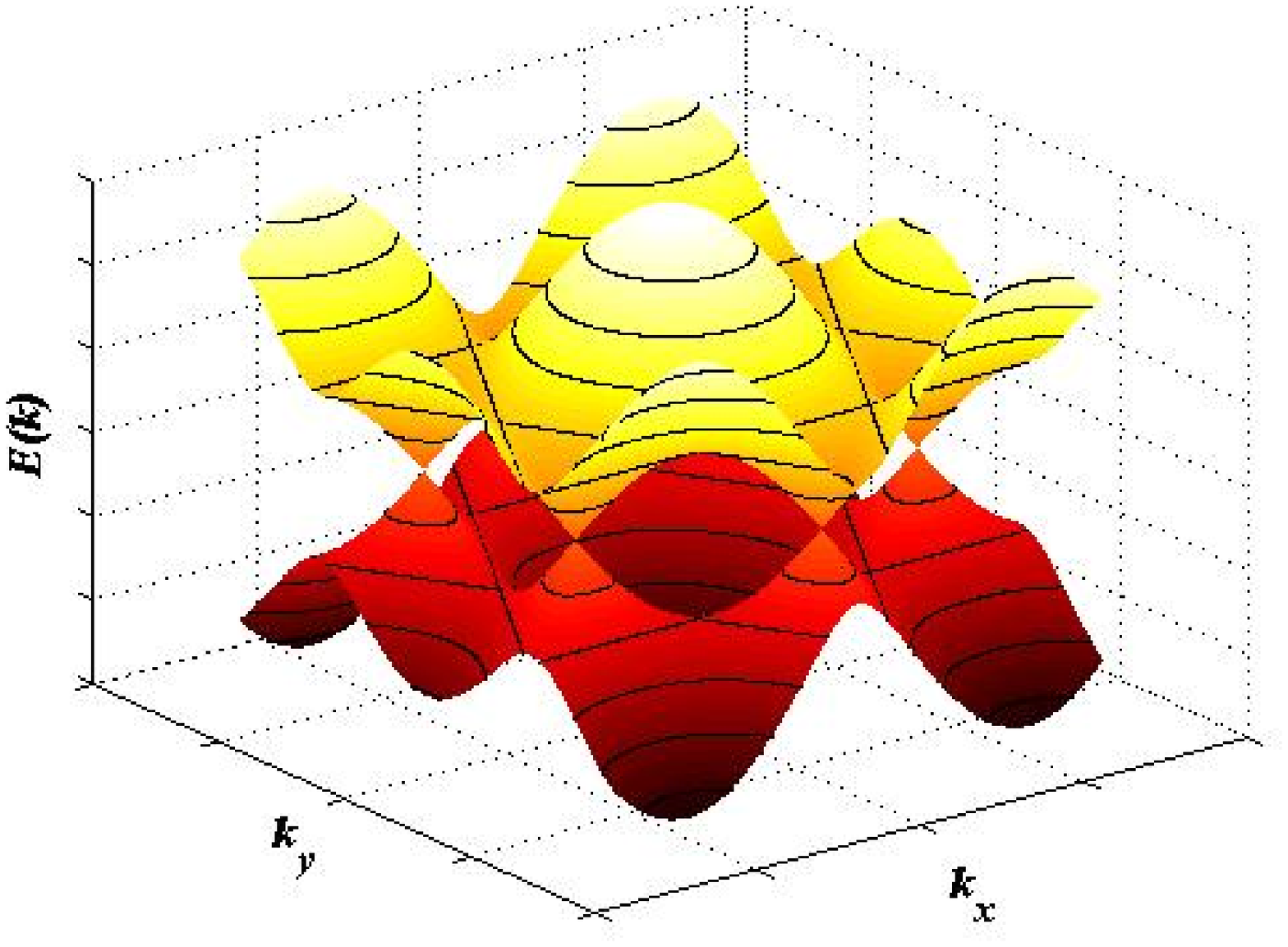}
\caption{Left: The honeycomb lattice is made of two
interpenetrating triangular lattices. Right: the dispersion
relation.}
    \label{lattice}
\end{center}
\end{figure}
Monolayer graphite - graphene- consists of a planar honeycomb
lattice of carbon atoms shown at the left hand side of Fig.
\ref{lattice}. In the graphene structure the in-plane $\sigma$
bonds are formed from 2$s$, 2$p_x$ and 2$p_y$ orbitals hybridized
in a $sp^2$ configuration, while the 2$p_z$ orbital, perpendicular
to the layer, builds up covalent bonds, similar to the ones in the
benzene molecule. The $\sigma$ bonds give rigidity to the
structure and the $\pi$ bonds give rise to the valence and
conduction bands. A usual tight binding analysis of the system
\cite{W47,SW58} leads to a special band structure shown at the
right hand side of Fig. \ref{lattice}. The Fermi surface of the
neutral system consists of six Fermi points (only two are
independent). A continuum model for the low energy excitations
around  the Fermi points (i=1,2) can be defined governed by the
Hamiltonian
\begin{equation}
{\cal H}_{0i}= i\hbar v_{\rm F} \int d^2 {\bf r} \bar{\Psi}_i({\bf
r}) [(-1)^i  \sigma_x \partial_x +  \sigma_y \partial_y ] \Psi_i
({\bf r})\;,\label{hamil}
\end{equation}
where $\sigma_{x,y}$ are the Pauli matrices, $v_{\rm F} = (3 t a
)/2 $, and $a=  1.4\AA$ is the distance between nearest carbon
atoms. The components of the two-dimensional wavefunction:
\begin{equation}
\Psi_i( {\bf r} )= \left( \begin{array}{c} \varphi_A ( {\bf r} )
\\  \varphi_B ( {\bf r }) \end{array} \right)_i
\label{2spinor}
\end{equation}
correspond to the amplitude of the wave function in each of the
two sublattices (A and B) which build up the honeycomb structure
and the subindex $i$ refers to the two Fermi points. It can be
shown that the wave function (\ref{2spinor}) transforms as a
bispinor in k-space and it acquires a phase of $\pi$ under a
$2\pi$ rotation.  This Berry  phase together with the helicity
conservation has important physical consequences as the absence of
backward scattering in the electronic transport what implies the
absence of weak localization.

The spinor structures around each of the two Fermi points remain
degenerate and independent in the absence of short range
interactions of disorder. As we will see, topological disorder
mixes the two Fermi points and in the modelling of them it will be
useful to combine the two bispinors into a four dimensional object
described by the effective Hamiltonian
\begin{eqnarray}
H_{D}=-iv_{F}\hbar(1 \otimes \sigma_{1} \partial_{x}+\tau^{3}
\otimes \sigma_{2} \partial_{y}) ,\label{hamil4}
\end{eqnarray}
where $\sigma$ and  $\tau$ matrices are Pauli matrices acting on
the sublattice and valley degree of freedom respectively.

\section{Modelling curvature in graphene}
When trying to take into account the curvature of the graphene
samples and its physical implications  there are two aspects to
consider: The first and of the most interesting questions is the
physical origin of the ripples and their dynamics.  The present
experimental situation seems to indicate that there is little
dependence of the corrugations with temperature --although
systematic studies have not been performed yet --. On the other
hand, bilayer structures also present corrugations but less
pronounced \cite{Metal07b}. These aspects concern the sigma bonds
of graphene and their elastic properties and involve energies of
the order of tens of eV. Although the elastic and mechanical
properties of carbon nanotubes have been extensively explored
\cite{SDD98}, very little can yet be found on the elastic
properties of plain graphene \cite{AIK08}.

A second aspect is that of the influence of the ripples on the
electronic properties of the samples. A possible  approach to this
problem is to assume that the samples are corrugated for whatever
reason and devise a model to study the implications of the
corrugations on the electronic properties. This is a sensible
procedure considering that  the electronic properties of graphene
are attached to the $\pi$ bonds and the associated processes
involve energies several orders of magnitude smaller than those
related to the elasticity of the $\sigma$ bonds.

Since the low energy excitations of graphene are well described by
the massless Dirac equation, a natural way to incorporate the
effect of the observed corrugations at low energies is couple the
Dirac equation to the given curved background. The main assumption
of this approach is that the elastic properties of the samples
--determined by the sigma bonds -- are decoupled from the (pi)
electron dynamics. The ripples can then be modelled by a fixed
metric space defined phenomenologically  from the observed
corrugations and the electronic properties of the system will be
found from the computation  the Green's function in the curved
space following the standard formalism set in gravitational
physics \cite{BD82,W96}.

Although the physical origin of the ripples is unknown, we could
distinguish two possibilities: those coming from the curvature in
the substrate \cite{ICCFW07} that can be modelled by a smooth
curved space and those observed in the free standing samples. The
last class needs to be modelled by including topological defects
\cite{N02}: disclinations and disclination dipoles (dislocations).
In what follows we will summarize the general formalism to include
a curved background and to extract the physical implications.

The dynamics of a massless Dirac spinor in a curved spacetime is
governed by the modified Dirac equation:
\begin{equation}
i\gamma^{\mu}({\bf r})\nabla_{\mu}\psi=0 \label{dircurv}
\end{equation}
The curved space $\gamma$ matrices depend on the point of the
space and can be computed from the  anticommutation relations
\begin{equation}
\{\gamma^{\mu}({\bf r}),\gamma^{\nu}({\bf r})\}=2g^{\mu\nu}({\bf
r}).\nonumber
\end{equation}
The covariant derivative operator is defined as
$$
\nabla_{\mu}=\partial_{\mu}-\Omega_{\mu}
$$
where $\Omega_{\mu}$ is the spin connection of the spinor field
that can be calculated using the tetrad formalism\cite{BD82}.

Once the metric of the curved space is known there is a standard
procedure to get the geometric factors that enter into the Dirac
equation. In the modelling of the graphene ripples, the metric can
be treated as a smooth perturbation of the flat surface and
physical results are obtained by a kind of perturbation theory.
Very often, the final result can be cast in the form of the flat
Dirac problem in the presence of an effective potential induced by
the curvature.

The electronic properties of the curved sample can be extracted
from the two point Green's function. The equation for the exact
propagator in the curved space-time is:
\begin{equation}
i\gamma^{\alpha}e_{\alpha}^{\
\mu}\left(\partial_{\mu}+\Omega_{\mu}\right)\
G(x,x')=\delta(x-x')(-g)^{-\frac{1}{2}},
\end{equation}
where $\Omega_\mu$ is the spin connection  and $\sqrt{-g}$ is the
determinant of the metric. From the propagator one can compute the
local density of states:
\begin{equation}
\rho(E,\textbf{r})=-\frac{1}{\pi}Im Tr
[G(E,\textbf{r},\textbf{r})\gamma^{0}],
\end{equation}
and other one-particle properties of the system as the electron
lifetime.

The described formalism can in principle be worked out for any
curved background. The practical difficulties to implement this
scheme are related with the choice of a realistic metric
parametrizing the graphene sheet and to the ability  to perform a
"weak field" expansion of the metric around the flat case. In the
next sections we will see the mechanism at work in two quite
general examples.

\section{Smooth ripples from the substrate}
The case of smooth curvature was studied in  reference
\cite{JCV07} where a general metric was considered that was
non-singular over the surface and asymptotically flat.  The
embedding of a two-dimensional surface with polar symmetry --for
simplicity-- in three-dimensional space is described in
cylindrical coordinates by a function $z(r)$ giving the height
with respect to the flat surface z=0, and parametrized by the
polar coordinates of its projection onto the z=0 plane. The metric
for this surface is readily obtained  by computing
\begin{equation}
dz^{2}=(\frac{dz}{dr})^{2}dr^{2}\equiv \alpha f(r)dr^{2},
\label{surface}
\end{equation}
 and substituting for the line element:
\begin{equation}
ds^{2}=dr^{2}+r^{2}d\theta^{2}+dz^{2}=\left(1+\alpha
f(r)\right)dr^{2}+r^{2}d\theta^{2}. \label{generalmetric}
\end{equation}
where $f(r)$ is a smooth function with the appropriate asymptotic
behavior and the parameter $\alpha$ controls the deviations from
flat space. The advantage of this approach is that it provides
already the perturbative parameter needed to compute the electron
propagator.

\begin{figure}
\begin{center}
\includegraphics[width=6cm]{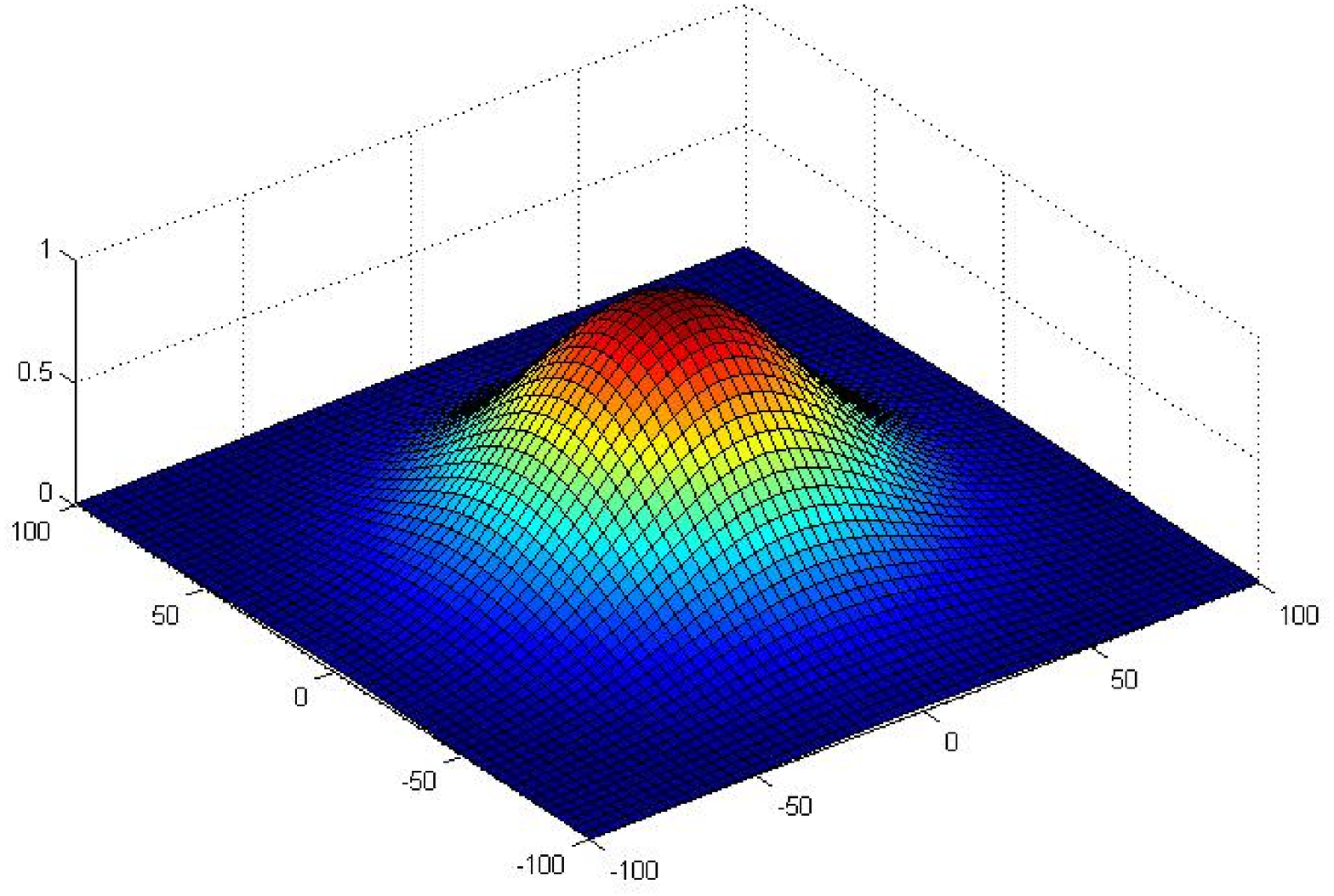}
\hspace{1.5cm}
\includegraphics[width=5cm]{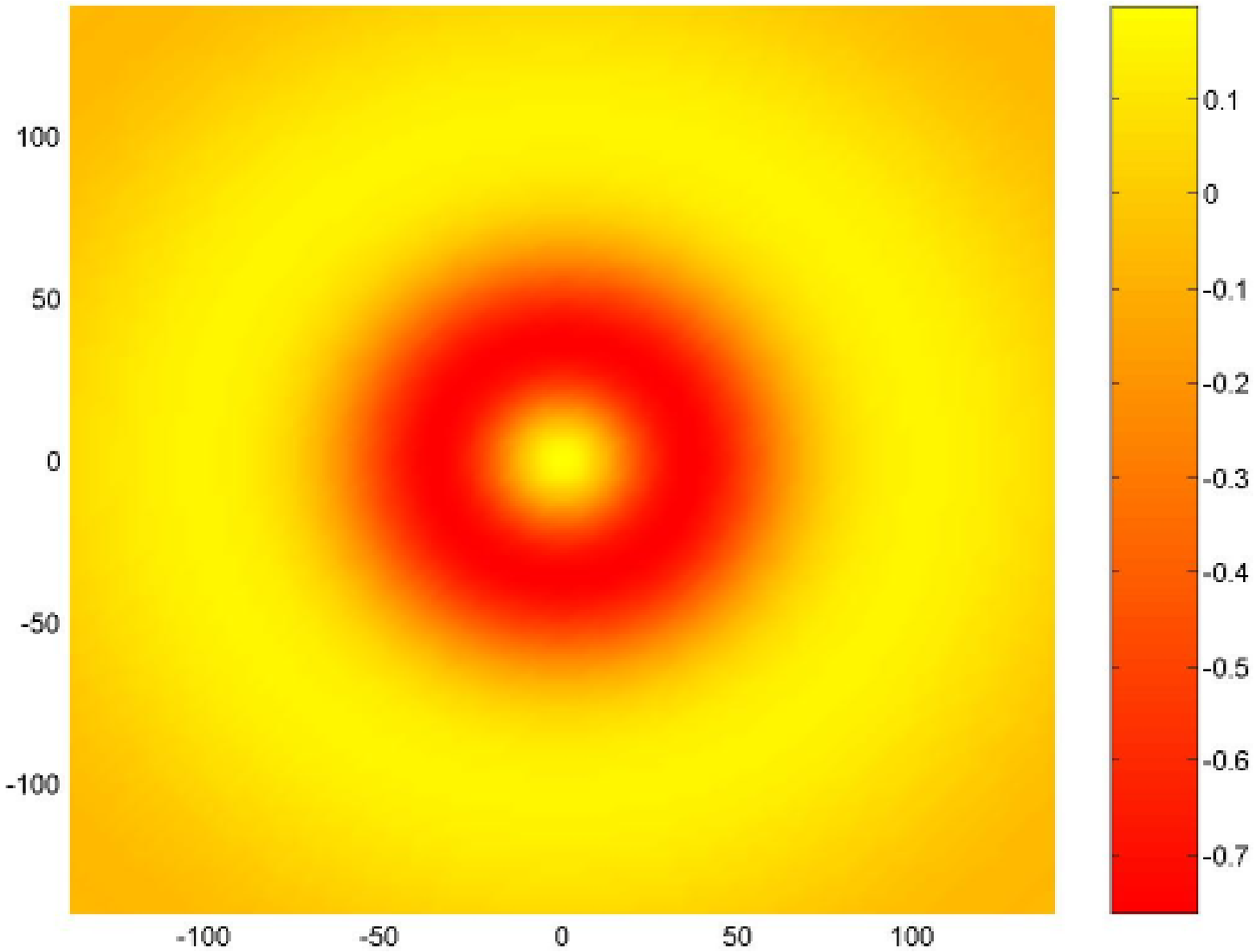}
\caption{Left: A smooth curved bump  in the graphene sheet. Right:
Effect of the curved bump shown in the left on the local density
of states. The color code is indicated in the figure. Darker
(lighter) areas represent negative (positive) corrections to the
density of states of the flat graphene sheet.} \label{gaussian}
\end{center}
\end{figure}

 A concrete example is provided by the gaussian shape shown at
the left hand side of fig. \ref{gaussian}. It models a smooth
protuberance fitting without singularities in the average flat
graphene sheet.
\begin{equation}
f(r)=4(r/b)^{2}\exp(-2r^{2}/b^{2}),
\end{equation}
This example was worked out in full detail in ref. \cite{JCV07}. A
comparison of the Dirac Hamiltonian in the plane (flat) in polar
coordinates
\begin{equation}
H_{flat}=\hbar v_F
\left(%
\begin{array}{cc}
  0 & \partial_r+i\frac{\partial_\theta}{r}+\frac{1}{2r} \\
  \partial_r-i\frac{\partial_\theta}{r}+\frac{1}{2r} & 0  \\
\end{array}%
\right) \label{Hflat}
\end{equation}
with the corresponding curved hamiltonian
\begin{equation}
H_{curved}=\hbar v_F
\left(%
\begin{array}{cc}
  0 & (1+\alpha f(r))^{-1/2}\partial_r+i\frac{\partial_\theta}{r}+A_\theta \\
  (1+\alpha f(r))^{-1/2}\partial_r-i\frac{\partial_\theta}{r}+A_\theta & 0  \\
\end{array}
\right), \label{Hcurved}
\end{equation}
\begin{equation}
A_\theta=\frac{\Omega_\theta}{2r}=\frac{1-(1+\alpha
f)^{-1/2}}{2r},
\end{equation}
allows to extract the two main features of the model. First we can
see that the curved bump produces an effective Fermi velocity
$\tilde{v}_{r}$ in the radial direction given by
\begin{equation}
\tilde{v}_r(r,\theta)=v_F(1+\alpha f(r))^{-1/2}.
\end{equation}
The effective Fermi velocity will always be smaller in magnitude
than the flat one. For a general curved surface described in polar
coordinates by an arbitrary function $z=z(r)$ it will be
\begin{equation}
v_r=\frac{v_0}{\sqrt{1+z'(r)^2}}.
\end{equation}
In a more general case we will have the two components of the
velocity changed but always to a smaller value irrespective of the
sign of the curvature at the given point.

The second feature that arises is an effective magnetic field
perpendicular to the graphene sheet  given by
\begin{equation}
B_z=-\frac{1}{r}\partial_r(rA_\theta)=\frac{1}{4r}\frac{\alpha
f'}{(1+\alpha f)^{3/2}}. \label{B}
\end{equation}

The electronic properties of the curved sample were computed from
the electron propagator to first  order in the small parameter
$\alpha$ that measures the deviation from flat space. In our
example, $\alpha=(A/b)^{2}$ is the (squared) height to length
ratio of the gaussian, so for typical ripples in graphene
$\alpha\approx 0.01$, since this ratio is of the order of 0.1
\cite{Metal07}. They are shown in the right hand side of Fig.
\ref{gaussian}. As a general feature  the curvature induces
inhomogeneities in the electronic density of states that could be
related to the observations reported in \cite{Y07}. It is worth
noticing that the contribution of the effective gauge field coming
from the spin connection to first order in perturbation theory
vanishes and all the corrections come from the determinant of the
metric and the curved gamma matrices. This is important when
trying to compare the present formalism with the tight binding
modelling of curvature where only the gauge field arises from the
modulation of the hopping parameter.

\section{Topological defects}
In the absence of a substrate or strain fields the only way to
have intrinsic curvature in the samples is by the presence of
topological defects \cite{N02}. In the case of the hexagonal
lattice of graphene disclinations form by replacing a hexagon of
the lattice by a n--sided polygon with $n\neq 6$. The most common
disclination in graphene is the pentagon that plays an important
role in the formation of fullerenes \cite{KHetal85}. Dislocations
are made of pentagon--heptagon pairs and they have been widely
studied in connection with the properties of carbon nanotubes
\cite{SDD98} and, more recently, in graphene
\cite{Aetal94,CB08,Cetal08,LJV08}. Observation of topological
defects in graphene have been reported in
\cite{Aetal01,Hetal04,DML04}.

Following the discovery of the fullerenes and nanotubes
topological defects have been modelled in the hexagonal lattice in
different ways. A very interesting approach related with the
theory of elasticity  is the gauge theory of the defects described
in refs. \cite{K89,KO99,OKP03,FCR06} or the metric formulation of
the theory of defects in solids set in \cite{KV92}. Most of these
approaches model the topological defects focussing on their
"holonomy" \cite{GGV92,GGV93,LC00,OK01,FMC06}and very little
effort has been devoted to the specific issue of the curvature in
these cases.  The curvature of the graphene surface in the
presence of a disclination in the continuum limit has a
delta-function singularity at the position of the defective ring.
In modelling spherical \cite{GGV92,GGV93} and quasi-spherical
\cite{KO06,PPO06} fullerenes curvature effects are included but
the curvature is averaged along the spherical surface. The more
difficult task of modelling several topological defects located at
arbitrary positions was undertaken in refs. \cite{CV07a,CV07b}. An
equal number of pentagon and heptagon defects was considered to
keep the samples flat in average. The  proposed metric was taken
from the cosmic string scenario:
\begin{equation}
ds^{2}=-dt^{2}+e^{-2\Lambda(x,y)}(dx^{2}+dy^{2}),\label{genmetric}
\end{equation}
where $$\Lambda(\textbf{r})=\sum^{N}_{i=1}4\mu_{i}\log(r_{i})$$
and
$$r_{i}=[(x-a_{i})^{2}+(y-b_{i})^{2}]^{1/2}.$$ This metric describes
the space-time around N parallel cosmic strings, located at the
points $(a_{i},b_{i})$. The parameters $\mu_{i}$ are related to
the angle defect or surplus $c_i$ by the relation
$c_{i}=1-4\mu_{i}$ in such manner that if $c_{i}<1 (>1)$ then
$\mu_{i}>0 (<0)$. Within the formalism described earlier it can be
seen that the Green's function is modified by the potential
\begin{equation}
V(\omega,\textbf{r})=2i\Lambda\gamma^{0}\partial_{0}+i\Lambda\gamma^{j}
\partial_{j}+\frac{i}{2}\gamma^{j}(\partial_{j}\Lambda).\label{effV}
\end{equation}
The computation of the corrections to the local density of states
for different positions of the defects showed   that pentagonal
(heptagonal) rings enhance (depress) the electron density, a
result that was obtained previously  \cite{TT94} with numerical
simulations. Numerical ab initio calculations show sharp resonant
peaks  in the LDOS at the tip apex of nanocones \cite{TT94,CR01}
that have been proposed for electronic applications in field
emission devices. Similar results have also been obtained
analytically in \cite{SV07}. A recent calculation of the minimal
conductivity in graphene with a random distribution of heptagon
and pentagon rings bases on this model has been performed in
\cite{CV08}.

\section{Conclusions and future}
From the present and similar analyses we can conclude that the
morphology of the graphene samples is correlated with the
electronic properties. In particular the presence of ripples,
irrespective of their origin, induces corrections to the density
of states and affects the transport properties. The spatial
variation of the Fermi velocity is a distinctive prediction of the
given geometric formalism.   We note that the effective Fermi
velocity is the fitting parameter used in most experiments
\cite{Y07} whose interpretation might change if the possibility of
a space-dependent velocity is considered.

Gauge fields are abundant in graphene and arise in many different
contexts. This is partially a proof of the robustness of the Dirac
formulation where only minimal coupling in the form of electronic
current-vector field are marginal interactions to be considered at
low energies \cite{GGV94}. They arise in various contexts and it
would be interesting to have a complete classification. In the
present context it is interesting to note the work of \cite{M07}
where  it was shown that, around each Fermi point, static strains
can mimic the effects of external electric and magnetic fields.

Future work should  address a better understanding of the
mechanisms leading to ripple formation and elastic properties of
graphene. There is also little work done on transport and
localization properties of graphene in the presence of topological
defects, a very important issue.

\vspace{1cm}

{\bf Acknowledgments}

This work is support by MEC (Spain) through grant
FIS2005-05478-C02-01 and by the European Union Contract 12881
(NEST) Ferrocarbon.

\bibliography{Vozmediano}

\end{document}